\documentclass{aipproc}
\layoutstyle{8x11double}

\usepackage{amssymb}

\begin{document}

\title[AGN Observations with MAGIC]{Scientific Highlights from Observations of Active Galactic Nuclei with the MAGIC Telescope
}

\author{Robert Wagner (for the MAGIC Collaboration)}{
  address={Max-Planck-Institut f\"ur Physik, D-80805 M\"unchen, Germany},
}

\begin{abstract}
Since 2004, the MAGIC $\gamma$-ray telescope has newly discovered
6 TeV blazars. The total set of 13 MAGIC-detected active galactic
nuclei includes well-studied objects at other wavelengths like
Markarian 501 and the giant radio galaxy M\,87, but also the
distant the flat-spectrum radio quasar 3C\,279, and the newly
discovered TeV $\gamma$-ray emitter S5\,0716+71. In addition, also
long-term and multi-wavelength studies on well-known TeV blazars
and systematic searches for new TeV blazars have been carried out.
Here we report selected highlights from recent MAGIC observations
of extragalactic TeV $\gamma$-ray sources, emphasizing the new
physics insights MAGIC was able to contribute.
\end{abstract}

\maketitle

\section{Introduction}
The study of very high energy (VHE, $E\gtrsim 70$ GeV)
$\gamma$-ray emission from active galactic nuclei (AGN) is one of
the major goals of ground-based $\gamma$-ray astronomy. The
sensitivity of the current imaging air Cerenkov telescopes (IACT)
enables studies of the blazar phenomenon, and in particular
advances in understanding both the origin of the VHE $\gamma$-rays
as well as the relations between photons of different energies
(from radio to VHE).

Except for the radio galaxy M\,87, all 23 currently known VHE
$\gamma$-ray emitting AGNs \cite{Wagner3} are high-frequency peaked
BL~Lac objects \cite{padovani},\footnote{See {\tt
http://www.mpp.mpg.de/$\sim$rwagner/sources/} for an up-to-date
source list.} a subclass of blazars characterized by a low
high-energy luminosity when compared with quasars, and a synchrotron peak in
the X-ray band. Their Spectral Energy Distribution (SED) is
characterized by a second peak at very high $\gamma$-ray energies. In
synchrotron-self-Compton (SSC) models it is assumed that the
observed $\gamma$-ray peak is due to the inverse-Compton (IC)
emission from the accelerated electrons up-scattering previously
produced synchrotron photons to high energies \cite{mar}. In
hadronic models, instead, interactions of a highly relativistic
jet outflow with ambient matter, proton-induced cascades, or
synchrotron radiation off protons, are responsible for the
high-energy photons. Another defining property of blazars is the
high variability of their emission ranging from radio to
$\gamma$-rays. For VHE $\gamma$-ray blazars, correlations between
X-ray and $\gamma$-ray emission have been found on time scales
ranging from $\approx 10$ minutes to days and months (see, e.g.,
\citet{fossati,lenain}), although the correlations have proven to
be rather complicated \cite{Wagner4}.

Here we present selected results on multi-wavelength campaigns
with MAGIC participation and for MAGIC blazar observations of
Mkn~501, M\,87 (February 2008), 1ES\,1011+496, S5\,0716+71, and
3C\,279.

\section{The MAGIC Telescope}
MAGIC \citep{magic04}, located on the Canary Island of La Palma (2200~m
a.s.l.), is currently the largest (17-m) single-dish IACT. Its energy range
spans from 50--60~GeV (trigger threshold at small zenith angles) up to tens of
TeV. MAGIC has a sensitivity of $\sim$~1.6\% of the Crab Nebula flux in 50
observing hours. Its energy resolution is about 30\% above 100 GeV and about
25\% from 200 GeV onwards. The MAGIC standard analysis chain is described,
e.g., in \citet{Crab}.  Observations during moderate moonshine enable a
substantially extended duty cycle \citep{jrico}, which is particularly
important for blazar observations. Parallel optical $R$-band observations are
performed by the Tuorla Blazar Monitoring Program and its KVA 35-cm telescope.
A second MAGIC telescope is being commissioned.  By stereoscopic observations,
the sensitivity of the two-telescope observatory will be further increased
considerably.

\section{Strong Flaring of Messier\,87 in February 2008}
\begin{figure}
 \resizebox{.95\textwidth}{!}
  {\includegraphics{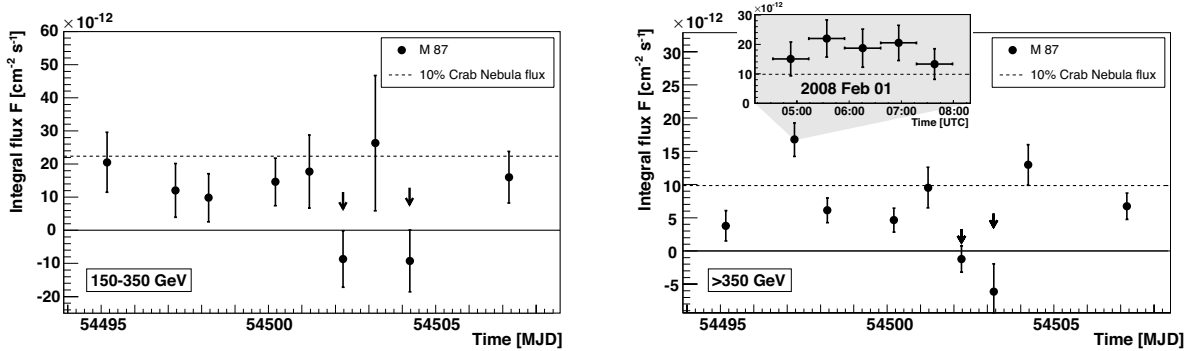}}
\caption{\label{fig:m87} The night-by-night light curve for M\,87
as measured from 2008 January 30 (MJD 54495) to 2008 February 11
(MJD 54507). Left: The flux in the energy bin $150 - 350$\,GeV,
being consistent with a constant emission. Right: Integral flux
above 350\,GeV; flux variations are apparent on variability
timescales down to 1 day. The inset shows the light curve above
350 GeV in a 40 min time binning for the night with the highest
flux (2008 February 1). The vertical arrows represent flux upper
limits (95\% c.l.) for the nights with negative excesses.}
\end{figure}
M\,87 is the only non-blazar radio galaxy known to emit VHE
$\gamma$-rays, and one of the best-studied extragalactic black-hole
systems. To enable long-term studies and assess the variability
timescales and the location of the VHE emission in M\,87, the
H.E.S.S., VERITAS, and MAGIC collaborations established a regular,
shared monitoring of M\,87 and agreed on mutual alerts in case of
a significant detection \cite{beilicke}. Results from the entire
monitoring campaign will appear in a separate paper. During the
MAGIC observations, a strong signal of $8\sigma$ significance was
found on 2008 February 1, triggering the other IACTs as well as
{\it Swift} observations. For the first time, we assessed the
energy spectrum below 250 GeV \cite{hdgsm87,MAGICM87}, which can
be described by a power law with a spectral index of $\Gamma=2.30
\pm 0.11_\mathrm{stat} \pm 0.20_\mathrm{syst}$. This relatively
hard VHE spectrum is unique among the extragalactic VHE
$\gamma$-ray sources, which show either curved or softer spectra.
We did not measure a high-energy cut-off, but found a marginal
spectral hardening, which may be interpreted as a similarity to
other blazars detected at VHE, where such hardening has often been
observed \cite{Wagner3}.

Our analysis revealed a variable (significance: 5.6\,$\sigma$)
night-to-night $\gamma$-ray flux above 350 GeV, while no
variability was found in the 150--350 GeV range
(Fig.~\ref{fig:m87}). We confirm the $E>730\,\mathrm{GeV}$
short-time variability of M\,87 reported by \citet{hessm87}. This
fastest variability $\Delta t$ observed so far in TeV $\gamma$-rays 
in M\,87 is on the order of or even below one day, restricting the
emission region to a size of $R\leq \Delta t\,c\,\delta = 2.6
\times 10^{15}\,\mathrm{cm} = 2.6\,\delta $ Schwarzschild radii (Doppler factor
$\delta$), and suggests the core of M\,87 rather than the brightest known knot
in the M\,87 jet, HST-1, as the origin of the TeV $\gamma$-rays.

There exists no lower limit on the size of HST-1, and thus the
flux variability cannot fully dismiss HST-1 as possible origin of
the TeV flux. During the MAGIC observations, however, HST-1 was at
a historically low X-ray flux level, whereas at the same time the
luminosity of the M\,87 core was at a historical maximum (D.
Harris, priv. comm.). This strongly supports the core of M\,87 as
the VHE $\gamma$-ray emission region.

\section{The July-2005 Flares of Mkn 501}
Mkn 501 ($z=0.034$) is known to be a strong and variable VHE
$\gamma$-ray emitter. MAGIC observed Mkn~501 for 24 nights during
six weeks in summer 2005 \cite{alb07a}. In two of these (one with
moon present), the recorded flux exceeded four times the
Crab-nebula flux, and revealed rapid flux changes with doubling
times as short as 3 minutes or less. For the first time, short
($\approx$~20 min) VHE $\gamma$-ray flares with a resolved time
structure could be used for detailed studies of particle
acceleration and cooling timescales.

Interestingly the flares in the two nights behaved differently:
While the 2005 June 30 flare is only visible in
250\,GeV--1.2\,TeV, the 2005 July 9 flare is apparent in all
energy bands (120~GeV to beyond 1.2~TeV). Additionally, under some
assumptions a time delay between the flare peak in different
energy bins ($E<0.25$~TeV and $E>1.2$~TeV) in the latter flare was
determined to be $4\pm1$~minutes \cite{alb07a}. While this
comparatively high value may be explained by the crude binned
analysis, the indicated error is clearly underestimated. The
reanalysis of these data in \cite{alb07b} resulted in a
much-improved estimate for the time-energy relation. At a
zero-delay probability of $P=0.026$, a marginal time delay of
$\tau_l = (0.030\pm0.012)\,\mathrm{s\,GeV}^{-1}$ towards higher
energies was found using two independent analyses, both exploiting
the full statistical power of the dataset (see \cite{alb07b,me08}
for details).

Several explanations for this delay have been considered up to
now:
\begin{enumerate}
\item Particles inside the emission region moving with constant
Doppler factor need some time to be accelerated to energies that
enable them to produce $\gamma$ rays with specific energies
\cite{alb07a}.
\item The $\gamma$-ray emission has been captured in the initial
acceleration phase of the relativistic blob in the jet, which at
any point in time radiates up to highest $\gamma$-ray energies
possible \cite{bwhdgs,bw08}.
\item An one-zone SSC model, which invokes a brief episode of
increased particle injection at low energies \cite{mas08}.
\item An energy-dependent speed of photons in vacuum \cite{mat}, as predicted
in some models of quantum gravity \cite{gac}. When assuming a simultaneous
emission of the $\gamma$-rays (of different energies) at the source, a lower
limit of $M_{\mathrm{QG}} > 0.21 \times 10^{18}$~GeV (95\% c.l.) can be
established \cite{alb07b}, which increases further if any delay towards higher
energies in the source is present. 
\end{enumerate}

\section{Blazars Detected during Optical Outbursts}
MAGIC has been performing target of opportunity observations upon
high optical states of known or potential VHE blazars. Up to now,
this strategy has proven very successful with the detection of Mkn
180, 1ES 1011+496, and recently S5\,0716+71 (see
\citet{OpticalHDGS} and references therein). The 18.7-h
observation of 1ES 1011+496 was triggered by an optical outburst
in March 2007, resulting in a 6.2-$\sigma$ detection at
$F_{>200\mathrm{GeV}} = (1.58\pm0.32)\times10^{-11}
\mathrm{cm}^{-2} \mathrm{s}^{-1}$ \cite{MAGIC1011}. An indication
for an optical--VHE correlation is given, in that in spring 2007
the VHE $\gamma$-ray flux is >40\% higher than in spring 2006,
where MAGIC observed the blazar as part of a systematic search for
VHE emission from a sample of X-ray bright
($F_{1\,\rm{keV}}$\,$>$\,$2\,\rm{\mu Jy}$) northern HBLs at
moderate ($z$\,$<$\,$0.3$) redshifts \cite{systHBL}.

In April 2008, KVA observed a high optical state of the blazar
S5~0716+71, triggering MAGIC observation, which resulted in a the
detection of a strong 6.8-$\sigma$ signal, corresponding to a flux
of $F_{>400\mathrm{GeV}} \approx 10^{-11} \mathrm{cm}^{-2}
\mathrm{s}^{-1}$. 
The MAGIC observation time was 2.6\,h. The source was also in a
high X-ray state \cite{atel1495giommi}.

The determination of the before-unknown redshifts of 1ES\,1011+496
($z=0.21$, \cite{MAGIC1011}) and S5\,0716+71 ($z=0.31$,
\cite{nps08}) makes these objects the third-most and second-most
distant TeV blazars after 3C\,279, respectively.

\section{Multi-Wavelength Campaigns}
For an advanced understanding of blazars, coordinated simultaneous
multi-frequency observations are essential, yielding SEDs spanning
over 15 orders of magnitude. MAGIC participated in a number of
multiwavelength-campaigns carried out on five Northern-hemisphere
blazars, which involved the X-ray instruments {\it Suzaku} and
{\it Swift}, the $\gamma$-ray telescope H.E.S.S. and MAGIC, and
other optical and radio telescopes.
\begin{itemize}
\item Mkn\,421 was detected in a campaign April 2006 in all
wavelengths (publication under preparation),
\item The observations of Mkn\,501 in July 2006 revealed the
lowest X-ray and VHE state ever observed. No variability in VHE
$\gamma$-rays was found, while an overall increase of about 50\%
during one day was seen in X-rays. A one-zone SSC model describes
this quiescent state of Mkn\,501 well \cite{mahaya}.
\item Also campaigns on 1ES\,1218+304 and 1H\,1426+428 have been
carried out, during both of which significant X-ray variability
has been observed. The VHE data are being analyzed.
\item The VHE emission of PG\,1553+113 showed no variability
during the first multi-wavelength campaign on this blazar in July
2006 \cite{reimer2,reimer}.
\item 1ES\,1959+650 showed VHE data among the lowest flux state
observed from this object, while at the same time a relatively
high optical and X-ray flux (both Swift/Suzaku) was found
\cite{tag}. The SED could be modeled using similar parameters as
needed for the SED measured in 2002, with a slightly more compact
source and a slightly lower magnetic field.
\end{itemize}
Further campaigns have been organized.

\section{Detection of the flat-spectrum radio quasar 3C\,279}
\begin{figure}
 \resizebox{\columnwidth}{!}
  {\includegraphics{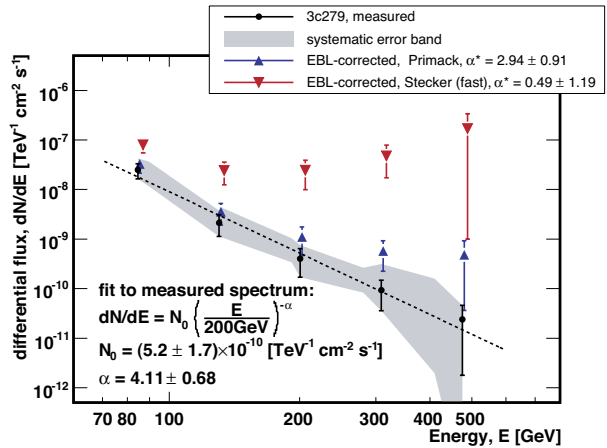}}
\caption{\label{fig:279} Differential energy spectrum of 3C~279.
The grey area includes the combined statistical ($1\sigma$) and
systematic errors. The dotted line shows compatibility of the
measured spectrum with a power law of photon index $\alpha=4.1$.
The triangles are measurements corrected on the basis of two
models for the EBL density (see text). 
}
\end{figure}
Observations of 3C~279, the brightest EGRET AGN \cite{wehrle},
during the WEBT multi-wavelength campaign \cite{webt} revealed a
5.77-$\sigma$ post-trial detection on 2006 February 23 supported
by a marginal signal on the preceding night
\cite{MAGICScience,Errando}. The overall probability for a
zero-flux lightcurve can be rejected on the 5.04-$\sigma$ level.
Simultaneous optical $R$-band observations, by the Tuorla
Observatory Blazar Monitoring Program revealed that during the
MAGIC observations, the $\gamma$-ray source was in a generally high
optical state, a factor of 2 above the long-term baseline flux,
but with no indication of short time-scale variability at visible
wavelengths. The observed VHE spectrum can be described by a power
law with a differential photon spectral index of
$\alpha=4.1\pm0.7_\mathrm{stat}\pm0.2_\mathrm{syst}$ between 75
and 500 GeV (Fig. \ref{fig:279}). The measured integrated flux
above 100~GeV on 23~February is $(5.15\pm 0.82_\mathrm{stat}\pm
1.5_\mathrm{syst}) \times 10^{-10}$ photons cm$^{-2}$ s$^{-1}$.

Located at $z=0.536$, VHE observations of such distant sources
were until recently impossible due to the expected strong
attenuation of $\gamma$ rays by the extragalactic background light
(EBL; see \cite{hauser} for a review), which influences the
observed spectrum and flux, resulting in an exponential decrease
with energy and a cutoff in the $\gamma$-ray spectrum. The observed
VHE spectrum is sensitive to the EBL between $0.2-2
\mu\mathrm{m}$. The reconstructed intrinsic spectrum is difficult
to reconcile with models predicting high EBL densities (e.g., the
fast-evolution model of \citet{stecker}), while low-level models,
e.g. \cite{primack,ahaebl}, are still viable. Assuming a maximum
intrinsic photon index of $\alpha^\ast = 1.5$, an upper EBL limit
is inferred, leaving a small allowed region for the EBL.

The results support, at higher redshift, the conclusion drawn from
earlier measurements \cite{ahaebl} that the observations of the
{\it Hubble Space Telescope} and {\it Spitzer} correctly estimate
most of the light sources in the Universe. The derived limits are
consistent with the EBL evolution corresponding to a maximum
star-formation rate at $z \geq 1$, as suggested by \cite{primack}
and similar models.

\vspace{-0.2cm}
\begin{theacknowledgments}
\vspace{-0.1cm}
MAGIC enjoys the excellent working conditions at the ORM in La
Palma and is supported by the German BMBF and MPG, the Italian
INFN, the Spanish MCI-NN, ETH research grant TH~34/04~3, the Polish MNiSzW
Grant N~N203 390834, and by the YIP of the Helmholtz Gemeinschaft.
The author acknowledges partial support by the DFG Cluster of Excellence
``Origin and Structure of the Universe''.
\end{theacknowledgments}

\bibliographystyle{aipprocl}

\begin{thebibliography}{9}
\bibitem{Wagner3} R. M. Wagner \newblock 2008, {\em Mon. Not. R. Astr. Soc.}, 385, 119.
\bibitem{padovani} P. Padovani, P. Giommi \newblock 1995, {\em Astrophys. J.}, 444, 567.
\bibitem{mar} L. Maraschi et al. \newblock 1992, {\em Astrophys. J.}, 397, L5.
\bibitem[Fossati et~al.(2008)]{fossati}  G. Fossati et al. \newblock 2008, {\em Astrophys. J.}, 677, 906.
\bibitem[Lenain(2008)]{lenain} J. P. Lenain (H.E.S.S. Collab.) \newblock this volume.
\bibitem{Wagner4} R. M. Wagner \newblock 2008, PoS(BLAZARS2008)013. 
\bibitem{magic04} C. Baixeras et~al. (MAGIC Collab.) \newblock 2004, {\em Nucl. Instrum. Meth.}, A518, 188.
\bibitem[Albert et~al.(2008)]{Crab} J. Albert et~al. (MAGIC Collab.) \newblock 2008, {\em Astrophys. J.}, 674, 1037.
\bibitem{jrico}       J. Albert et~al. (MAGIC Collab.) \newblock 2007, preprint: arXiv:astro-ph/0702475.
\bibitem{beilicke}    M. Beilicke, M. Hui, D. Mazin, M. Raue, R. M. Wagner, S. Wagner (H.E.S.S., MAGIC, VERITAS Collab.) \newblock this volume.
\bibitem{hdgsm87}     D. Mazin, D. Tescaro, R. M. Wagner et al. (MAGIC Collab.) \newblock this volume.
\bibitem{MAGICM87}    J. Albert et al. (MAGIC Collab.) \newblock 2008, {\em Astrophys. J.}, 685, L23.
\bibitem[Aharonian et~al.(2006)]{hessm87}     F. Aharonian et al. (H.E.S.S. Collab.) \newblock 2006, {\em Science}, 314, 1424.
\bibitem{alb07a}      J. Albert et al. (MAGIC Collab.) \newblock 2007, {\em Astrophys. J.}, 669, 862.
\bibitem{alb07b}      J. Albert et al. (MAGIC Collab.) \newblock 2008, {\em Phys. Lett. B}, 668, 253. 
\bibitem{me08}        M. Martinez, M. Errando \newblock 2008, {\em Astropart. Phys.} submitted, preprint: arXiv:0803.2120 [astro-ph].
\bibitem{bwhdgs}      W. Bednarek, R. M. Wagner \newblock this volume.
\bibitem{bw08}        W. Bednarek, R. M. Wagner \newblock 2008, {\em A\&A}, 486, 679.
\bibitem{mas08}       A. Mastichiadis, K. Moraitis \newblock this volume.
\bibitem{mat}	      D. Mattingly \newblock 2005, {\em Living Rev. Rel.} 8, 5.
\bibitem{gac}         G. Amelino-Camelia, \newblock 2008, to appear in {\em Living Rev. Rel.}, preprint: arXiv:0806.0339 [gr-qc].
\bibitem{mahaya}      M. Hayashida et al. \newblock 2007, {\em in: Proc. 30th Int. Cosmic Ray Conf.}, M\'{e}rida, M\'{e}xico.
\bibitem{reimer2}     A. Reimer, O. Reimer, G. Madejski, D. Dorner \newblock this volume.
\bibitem{reimer}      A. Reimer, L. Costamante, G. Madejski, O. Reimer, D. Dorner \newblock 2008, {\em Astrophys. J.}, 682, 775.
\bibitem{tag}         G. Tagliaferri et al. (MAGIC Collab.) \newblock 2008, {\em Astrophys. J.}, 679, 1029.
\bibitem[Lindfors \& Mazin(2008)]{OpticalHDGS} E. Lindfors, D. Mazin (MAGIC Collab.) \newblock this volume.
\bibitem{MAGIC1011} J. Albert et al. (MAGIC Collab.) \newblock 2007, {\em Astrophys. J.}, 667, L21.
\bibitem{systHBL} J. Albert et al. (MAGIC Collab.) \newblock 2008, {\em Astrophys. J.}, 681, 944.
\bibitem{atel1495giommi} P. Giommi et al. \newblock 2008, {\em Astronomer's Telegram}, 1495, 1.
\bibitem{nps08} K. Nilsson, T. Pursimo, A. Sillanp\"a\"a, L. O. Takalo, E. Lindfors \newblock 2008, {\em Astron. Astrophys.}, 487, L29.
\bibitem{wehrle} A. E. Wehrle et al. \newblock 1998, {\em Astrophys. J.}, 497, 178.
\bibitem{webt} M. B\"ottcher et al. \newblock 2008, {\em Astrophys. J.}, 670, 968.
\bibitem{MAGICScience} J. Albert et al. (MAGIC Collab.) \newblock 2008, {\em Science}, 320, 1752.
\bibitem{Errando} M. Errando et al. (MAGIC Collab.) \newblock this volume.
\bibitem{hauser} M.~G.~Hauser, E.~Dwek \newblock 2001, {\em Ann. Rev. Astron. Astrophys.}, 39, 307.
\bibitem[Stecker et~al.(2007)]{stecker} F. W. Stecker et al. \newblock 2007, {\em Astrophys. J.}, 667, L29.
\bibitem{primack} J. R. Primack et al. \newblock 2005, {\em AIP Conf. Proc.}, 745, 23.
\bibitem{ahaebl} F. Aharonian et al. (H.E.S.S. Collab.) \newblock 2006, {\em Nature}, 440, 1018.
\end{thebibliography}

\end{document}